\documentclass[11pt,twoside]{article}

\begin{document}
\title{\bf Teleportation using continuous variable quantum cloning machine}
\author{Satyabrata Adhikari\thanks{Corresponding
Author:Satyabrata Adhikari,E-Mail:satyabrata@bose.res.in}\\
S. N. Bose National Centre for Basic Sciences, Salt Lake, Kolkata
700 098, India}
 \maketitle
 \begin{abstract}
 We show that an unknown quantum state in phase space can be
 teleported via three-mode entanglement generated by continuous
 variable quantum cloning machine (transformation). Further, proceeding with
 our teleportation protocol we are able to improve the
 fidelity of teleportation obtained by Loock et.al. [Phys.Rev.Lett. 84, 3482(2000)].
 Also we study here the entanglement between the two output copies from cloning
 machine.
 \end{abstract}
Quantum information theory was initiated with the discrete quantum
variables but now a days it has been extended to the domain of
continuous quantum variables. The possible reason for this may be
the continuous spectrum systems are experimentally simpler to
manipulate than their discrete counterparts in order to process
quantum information. The second reason to switch over from the
discrete quantum variables to the continuous variables is the
$\textit{Unconditionalness}$, which is one of the valuable feature
of quantum optical implementations based upon continuous
variables. But this valuable feature give a price in terms of the
quality of the entanglement. The entanglement and the
entanglement-based quantum protocol is always imperfect and the
degree of imperfection depends on the amount of squeezing of the
laser light involved. Good performance of the entanglement-based
quantum protocol can be achieved for large squeezing (about 10 dB
\cite{wu}) which is technologically demanding also. In this
letter, our aim is two fold: First, we consider the continuous
variable cloning transformation and study the entanglement between
the two output copies by covariance matrix approach. Second, We
will show that the three-mode entanglement (two copy mode and one
ancilla mode) generated by cloning procedure can be used to
teleport a single arbitrary mode with optimum fidelity $F^{opt}=
[\frac{1+3e^{-4r}+2e^{-6r}+2e^{-2r}}{1+2e^{-4r}}]^{\frac{-1}{2}}$.
We will observe that the optimum fidelity $F^{opt}$  obtained by
our protocol is greater than or equal to the optimum fidelity
$F'_{opt}=
[(1+e^{-2r})(1+\frac{3}{(2e^{2r}+e^{-2r})})]^\frac{-1}{2} $
obtained by Loock and Braunstein \cite{loock}.\\\\
Quantum cloning: An arbitrary quantum state cannot be cloned
perfectly but it does not rule out the approximate cloning
\cite{wz,bh}. There are numerous works in the literature on
quantum cloning in discrete variables as well as in continuous
variables
\cite{gm,bbhb,dg,sh,cerf1,cerf2,lc,bruss1,pati,levenson,mor,zanardi,gross,braunstein2,cerf3,bruss2,scarani,adhikari}.
The experimental realization of the Gaussian cloning machine with
three NOPAs has been discussed in \cite{ariano} but this cloning
device is approximate and the desired cloning transformation is
achieved only in the limit of infinite squeezing. Later this
problem was solved by Fiurasek \cite{fiurasek} and almost in the
same time by Braunstein et.al. \cite{braunstein1}. They showed
that such problems are avoided if sequence of beam splitters and a
phase-insensitive linear
amplifier are used to realize the cloning transformation.\\
The $1\rightarrow2$ continuous variable cloning transformation can
be realized by two canonical transformations (i) two mode
Bogoliubov transformation for phase-insensitive amplifier with
gain 2 applied on the input mode $\hat{a}_{0}$ and the ancilla
mode $\hat{a}_{z}$ and (ii) phase free 50:50 beam splitter
operation operated on the output mode $\hat{a}'_{0}$ from the
amplifier and on the blank mode $\hat{a}_{1}$
\begin{eqnarray}
\hat{a}_{0}^{'}=\sqrt{2}\hat{a}_{0}+\hat{a}_{z}^{\dagger},~~~~~\hat{a}_{z}^{'}=\hat{a}_{0}^{\dagger}+\sqrt{2}\hat{a}_{z},{}\nonumber\\
\hat{a}_{0}^{''}=\frac{(\hat{a}_{0}^{'}+\hat{a}_{1})}{\sqrt{2}},~~~~~\hat{a}_{1}^{''}=\frac{(\hat{a}_{0}^{'}-\hat{a}_{1})}{\sqrt{2}}
\end{eqnarray}
where $\hat{a}_{k}=\frac{(\hat{x}_{k}+i\hat{p}_{k})}{\sqrt{2}}$
and
$\hat{a}_{k}^{\dagger}=\frac{(\hat{x}_{k}-i\hat{p}_{k})}{\sqrt{2}}$
denote the annihilation and creation operators for mode k.
$\hat{a}_{0}$
is the input mode, $\hat{a}_{1}$ is the blank mode on which the information is to be copied and $\hat{a}_{z}$ is the ancilla mode.\\
In terms of position and momentum operators, the transformation
for any input mode reduces to
\begin{eqnarray}
\hat{x}_{0}^{''}=\hat{x}_{0}+\frac{\hat{x}_{1}}{\sqrt{2}}+\frac{\hat{x}_{z}}{\sqrt{2}},
~~~~~\hat{p}_{0}^{''}=\hat{p}_{0}+\frac{\hat{p}_{1}}{\sqrt{2}}-\frac{\hat{p}_{z}}{\sqrt{2}},{}\nonumber\\
\hat{x}_{1}^{''}=\hat{x}_{0}-\frac{\hat{x}_{1}}{\sqrt{2}}+\frac{\hat{x}_{z}}{\sqrt{2}},
~~~~~\hat{p}_{1}^{''}=\hat{p}_{0}-\frac{\hat{p}_{1}}{\sqrt{2}}-\frac{\hat{p}_{z}}{\sqrt{2}},{}\nonumber\\
\hat{x}_{z}^{'}=\hat{x}_{0}+\sqrt{2}\hat{x}_{z},~~~~~
\hat{p}_{z}^{'}=-\hat{p}_{0}+\sqrt{2}\hat{p}_{z}
\end{eqnarray}
If we assume, without loss of generality, that the input and blank
mode are squeezed in 'x' with squeezing parameter $r_{0}$ and
$r_{1}$ respectively and the ancilla mode is squeezed in 'p' with
squeezing parameter $r_{z}$ then the cloning transformation for
single mode squeezed state input in the Heisenberg picture is
given by
\begin{eqnarray}
\hat{x}_{0}^{''}=\hat{x}_{0}^{(0)}e^{-r_{0}}+\frac{\hat{x}_{1}^{(0)}e^{-r_{1}}}{\sqrt{2}}+\frac{\hat{x}_{z}^{(0)}e^{r_{z}}}{\sqrt{2}},
~~~~~\hat{p}_{0}^{''}=\hat{p}_{0}^{(0)}e^{r_{0}}+\frac{\hat{p}_{1}^{(0)}e^{r_{1}}}{\sqrt{2}}-\frac{\hat{p}_{z}^{(0)}e^{-r_{z}}}{\sqrt{2}},{}\nonumber\\
\hat{x}_{1}^{''}=\hat{x}_{0}^{(0)}e^{-r_{0}}-\frac{\hat{x}_{1}^{(0)}e^{-r_{1}}}{\sqrt{2}}+\frac{\hat{x}_{z}^{(0)}e^{r_{z}}}{\sqrt{2}},
~~~~~\hat{p}_{1}^{''}=\hat{p}_{0}^{(0)}e^{r_{0}}-\frac{\hat{p}_{1}^{(0)}e^{r_{1}}}{\sqrt{2}}-\frac{\hat{p}_{z}^{(0)}e^{-r_{z}}}{\sqrt{2}},{}\nonumber\\
\hat{x}_{z}^{'}=\hat{x}_{0}^{(0)}e^{-r_{0}}+\sqrt{2}\hat{x}_{z}^{(0)}e^{r_{z}},~~~~~
\hat{p}_{z}^{'}=-\hat{p}_{0}^{(0)}e^{r_{0}}+\sqrt{2}\hat{p}_{z}^{(0)}e^{-r_{z}}
\end{eqnarray}
After tracing out the ancilla mode and putting $r_{z}=0$, the
second moment correlation matrix (Covariance matrix) of the two
clone mode is given by
\begin{eqnarray}
\sigma_{12} = \left(\begin{matrix}{s & 0 & t & 0 \cr 0 & u & 0 & v
\cr t & 0 & s & 0 \cr 0 & v & 0 & u}\end{matrix}\right)
\label{twoancilla}
\end{eqnarray}
where
\begin{eqnarray}
s=\frac{1}{4}(e^{-2r_{0}}+\frac{e^{-2r_{1}}}{2}+\frac{1}{2}){}\nonumber\\
t=\frac{1}{4}(e^{-2r_{0}}-\frac{e^{-2r_{1}}}{2}+\frac{1}{2}){}\nonumber\\
u=\frac{1}{4}(e^{2r_{0}}+\frac{e^{2r_{1}}}{2}+\frac{1}{2}){}\nonumber\\
v=\frac{1}{4}(e^{2r_{0}}-\frac{e^{2r_{1}}}{2}+\frac{1}{2})
\end{eqnarray}
The fidelity of the two clones can be evaluated through the
relation \cite{olivares}
\begin{eqnarray}
F^{clone} = \frac{1}{\sqrt{\mathrm{Det}[\sigma_{in} +
\sigma_{out}]+\delta}-\sqrt{\delta}} \label{fidel0}
\end{eqnarray}
where $\delta =
4(\mathrm{Det}[\sigma_{in}]-1/4)(\mathrm{Det}[\sigma_{out}]-1/4)$.\\
The input state and the two output cloned states are described by
the covariance matrices
$\sigma_{in}=\left(\begin{matrix}{\frac{e^{-2r_{0}}}{4} & 0 \cr 0
& \frac{e^{2r_{0}}}{4}}\end{matrix}\right) \label{twoancilla}$ and
$\sigma_{1}^{clone} =\sigma_{2}^{clone} = \left(\begin{matrix}{s
&0\cr 0 & u}\end{matrix}\right)\label{clone0}$.\\
After some simple algebra, the fidelity of cloning is calculated
to be
\begin{eqnarray}
F^{clone} =
\frac{16}{\sqrt{434+71(e^{2r}+e^{-2r})}-\sqrt{18-9(e^{2r}+e^{-2r})}}
\end{eqnarray}
where $r_{0}=r_{1}=r$.\\
If the single mode coherent state $(i.e.~ \textrm{the mode with
squeezing parameter}~  r=0)$ is given to the
cloning machine as input then the fidelity of cloning is found to be $\frac{2}{3}$.\\
Now we investigate the qualification and quantification of
entanglement exist (if any) between the two clone modes. The
positivity of the partially transposed state (PPT) criterion is
necessary and sufficient for the separability of two-mode Gaussian
states \cite{simon,duan}. Further, J.Laurat et.al. \cite{laurat}
showed that the PPT criterion and an inequality satisfied by the
smallest symplectic eigen value $\tilde{\nu}_{-}$ of the partially
transposed state are equivalent. Therefore, the two mode Gaussian
state is entangled if the smallest symplectic eigenvalue
$\tilde{\nu}_{-}$ of the partially transposed state is less than
one i.e. $\tilde{\nu}_{-}<1$ provided that the eigen values of the
covariance matrix is positive. The eigen values of the covariance
matrix $\sigma_{12}$ is given by $\lambda_{1}= s+t,\lambda_{2}=
s-t, \lambda_{3}= u+v, \lambda_{4}= u-v$. We can now easily verify
that all the four eigen values of  $\sigma_{12}$ are positive for
any non-negative squeezing parameter and hence the state
represented by the covariance matrix $\sigma_{12}$ is physical. In
this work, we use the logarithmic negativity $E_{N}$ defined by
$E_{N}= max[0,-log_{2}(\tilde{\nu}_{-})]$ to quantify the
entanglement.\\
The smallest symplectic eigenvalues $\tilde{\nu}_{-}$ of the
partial transposed state $\tilde{\sigma}_{12}$ read
\begin{eqnarray}
\tilde{\nu}_{-}=\sqrt{\frac{\tilde{\Delta}(\sigma_{12})-\sqrt{\tilde{\Delta}(\sigma_{12})^{2}-4Det\sigma_{12}}}{2}}{}\nonumber\\
=\sqrt{(s+t)(u-v)}=\frac{1}{4}\sqrt{(2e^{-2r_{0}}+1)e^{2r_{1}}}
\end{eqnarray}
where $\tilde{\Delta}(\sigma_{12})=\Delta(\tilde{\sigma}_{12})=2(su-tv)$ and $Det(\sigma_{12})=(s^{2}-t^{2})(u^{2}-v^{2})$.\\
Now we consider two cases:\\
Case-I: If $r_{1}=0$ and $r_{0}\rightarrow\infty$ then
$\tilde{\nu}_{-}=\frac{1}{4}<1$. Therefore, if the blank mode is
prepared in a vacuum mode (coherent state) and for the large
squeezing of the input mode, the two clone modes are entangled.
The amount of entanglement is given by $E_{N}= log_{2}(4)$.  \\
Case-II: If $r_{0}=r_{1}=r$ then
$\tilde{\nu}_{-}=\frac{\sqrt{2+e^{2r}}}{4}$. The amount of
entanglement is given by $E_{N}=
-log_{2}(\frac{\sqrt{2+e^{2r}}}{4})$. From the figure (1), it is
clear that the two clone modes are entangled only when the
squeezing parameter $r$ takes the value lesser than 1.32.\\\\
Quantum teleportation: The idea of teleportation was first
introduced by Bennett et.al. \cite{bennett} and this ingenious
concept is about the transmission and reconstruction of the state
of a quantum system over arbitrary distances
\cite{bouwm,braunstein3}. The teleportation of continuous quantum
variables such as position and momentum of a particle, as first
proposed by Vaidman \cite{vaidman} relies on the entanglement of
the states in the original EPR paradox \cite{einstein}. In quantum
optical terms, the observables analogous to the two conjugate
variables position and momentum of a particle are the quadratures
of a single mode of the electromagnetic field. Continuous variable
quantum teleportation of arbitrary coherent states has been
realized experimentally with bipartite entanglement built from two
single-mode squeezed vacuum states combined at a beam splitter
\cite{furusawa}.\\
Loock and Braunstein \cite{loock} in their work showed that the
twice application of beam splitter operations known as tritter
\cite{braunstein4} defined by $\hat{T}_{123}\equiv
\hat{B}_{23}(\pi/4)\hat{B}_{12}(cos^{-1}\frac{1}{\sqrt{3}})$ to a
zero-momentum eigenstate in mode 1 and a pair of zero-position
eigenstates in modes 2 and 3 yields a GHZ-like state. Further they
showed that using the generated tripartite entanglement as a
quantum channel and with the help of classical communications, one
can teleport an unknown quantum state with optimum fidelity
$F'_{opt}=
[(1+e^{-2r})(1+\frac{3}{(2e^{2r}+e^{-2r})})]^\frac{-1}{2} $. They
found out that the perfect teleportation can be achieved for
infinite squeezing while if the squeezing parameters takes the
value zero then the optimal fidelity for teleportation achieved is
0.5. In this letter, we will show that the three mode entanglement
generated using the cloning transformation can be used to modify
the fidelity of teleportation $F'_{opt}$ .\\
Now we are in a position to discuss the teleportation of an
unknown quantum state through cloning procedure. Our teleportation
protocol can be illustrated in the following way: Firstly, Alice
use the cloning machine (1) to make two copies of a single mode
squeezed state $\hat{a}_{0}$. After the completion of copying
procedure, the entanglement between the two copy modes
$\hat{a}_{0}^{''}$ and $\hat{a}_{1}^{''}$ and the ancilla mode
$\hat{a}_{z}^{'}$ is generated. Alice then keeps one copy mode say
$\hat{a}_{0}^{''}$ and ancilla mode $\hat{a}_{z}^{'}$ with herself
and send another copy mode $\hat{a}_{1}^{''}$ to her distant
partner Bob. We want to use these three-mode (two copy mode and
one ancilla mode) entanglement as a quantum resource to teleport
an unknown quantum state. Alice now possesses two modes
$(\hat{a}_{0}^{''}$ and $\hat{a}_{z}^{'})$ and Bob holds the mode
($\hat{a}_{1}^{''}$). To teleport an arbitrary input mode
described in the phase-space
$\hat{a}^{in}=(\hat{x}_{in},\hat{p}_{in})$, Alice combines this
input mode with the copy mode
$\hat{a}_{0}^{''}=(\hat{x}_{0}^{''},\hat{p}_{0}^{''})$:
$\hat{x}_{m}= \frac{\hat{x}_{in}-\hat{x}_{0}^{''}}{\sqrt{2}}$,
$\hat{p}_{n}= \frac{\hat{p}_{in}+\hat{p}_{0}^{''}}{\sqrt{2}}$.
Bob's mode $\hat{a}_{1}^{''}=(\hat{x}_{1}^{''},\hat{p}_{1}^{''})$
and Alice's ancilla mode
$\hat{a}_{z}^{'}=(\hat{x}_{z}^{'},\hat{p}_{z}^{'})$ read as
\begin{eqnarray}
\hat{x}_{1}^{''}=\hat{x}_{in}-(\hat{x}_{0}^{''}-\hat{x}_{1}^{''})-\sqrt{2}\hat{x}_{m}{}\nonumber\\
\hat{p}_{1}^{''}=\hat{p}_{in}+(\hat{p}_{0}^{''}+\hat{p}_{1}^{''}+g^{(3)}\hat{p}_{z}^{'})-\sqrt{2}\hat{p}_{n}-g^{(3)}\hat{p}_{z}^{'}{}\nonumber\\
\hat{x}_{z}^{'}=\hat{x}_{in}-(\hat{x}_{0}^{''}-\hat{x}_{z}^{'})-\sqrt{2}\hat{x}_{m}{}\nonumber\\
\hat{p}_{z}^{'}=\hat{p}_{in}-(\hat{p}_{0}^{''}+g^{(3)}\hat{p}_{1}^{''}+\hat{p}_{z}^{'})-\sqrt{2}\hat{p}_{n}-g^{(3)}\hat{p}_{1}^{''}{}\nonumber\\
\end{eqnarray}
Where $g^{(3)}$ denotes the gain and its value which maximizes the
fidelity of teleportation will be determined later.\\
Alice then make measurements on the modes
$(\hat{x}_{m},\hat{p}_{n})$ and
$(\hat{x}_{z}^{'},\hat{p}_{z}^{'})$ and thereafter send the
measured values $(x_{m},p_{n})$ for $(\hat{x}_{m},\hat{p}_{n})$
and $(x_{z}^{'},p_{z}^{'})$ for $({x}_{z}^{'},{p}_{z}^{'})$ to Bob
through a classical channel. Then Bob displace his mode according
to the Alice's measured values. After displacement the teleported
mode in the Bob's side becomes
\begin{eqnarray}
\hat{x}_{tel}=\hat{x}_{in}-\sqrt{2}x_{1}^{(0)}e^{-r_{1}}{}\nonumber\\
\hat{p}_{tel}=\hat{p}_{in}+[(2-g^{(3)})\hat{p}_{0}^{(0)}e^{r_{0}}+\sqrt{2}(g^{(3)}-1)p_{2}^{(0)}e^{-r_{z}}]
\end{eqnarray}
Now to see the efficiency of our teleportation scheme, we have to
calculate the fidelity of teleportation. It is given by the
formula \cite{braunstein5}
\begin{eqnarray}
F= \pi Q_{tel}(x_{in}+ip_{in}){}\nonumber\\
= \frac{1}{2\sqrt{\sigma_{x}\sigma_{p}}}~~
exp[-(1-g)^{2}(\frac{x_{in}^{2}}{2\sigma_{x}}+\frac{p_{in}^{2}}{2\sigma_{p}})]
\end{eqnarray}
where g is the gain and $\sigma_{x}$ and $\sigma_{p}$ are the
variances of the Q function of the teleported mode for the
corresponding quadratures.\\
For g=1, it becomes
\begin{eqnarray}
F= \frac{1}{2\sqrt{\sigma_{x}\sigma_{p}}}
\end{eqnarray}
When $r_{0}=r_{1}=r_{z}=r$, the teleportation fidelity attains its
optimum value for
$g^{(3)}=\frac{2(e^{2r}+e^{-2r})}{(e^{2r}+2e^{-2r})}$. Therefore,
the optimum fidelity is given by
\begin{eqnarray}
F^{opt}=
[\frac{1+3e^{-4r}+2e^{-6r}+2e^{-2r}}{1+2e^{-4r}}]^{\frac{-1}{2}}
\end{eqnarray}
(i) If $r\rightarrow\infty$, then $g^{(3)}=2$ and $F^{opt}=1$.
Therefore, for infinite squeezing the optimum teleportation
fidelity goes to
unity and hence perfect teleportation is achieved.\\
(ii) If $r=0$ then  $g^{(3)}=\frac{4}{3}$ and
$F^{opt}=\frac{\sqrt{3}}{2\sqrt{2}}>\frac{1}{2}$. Note that in
this case the optimum fidelity $F^{opt}$ overtakes the optimum
fidelity $F'_{opt}$ obtained by Loock et.al. In figure(2), the
solid line represents the optimum fidelity $F^{opt}$ obtained in
our protocol and the dotted line represents the optimum fidelity
$F'_{opt}$ obtained in Loock's et.al. protocol. Therefore, with
the help of figure (2) we showed that the line of the optimum
fidelity $F^{opt}$ never go down the line of the optimum fidelity
$F'_{opt}$.\\
In summary, we showed that one can improve the fidelity of
teleportation of a single mode squeezed state to a certain extent
if continuous variable cloning machine is used to prepare
three-mode entangled state instead of tritter.

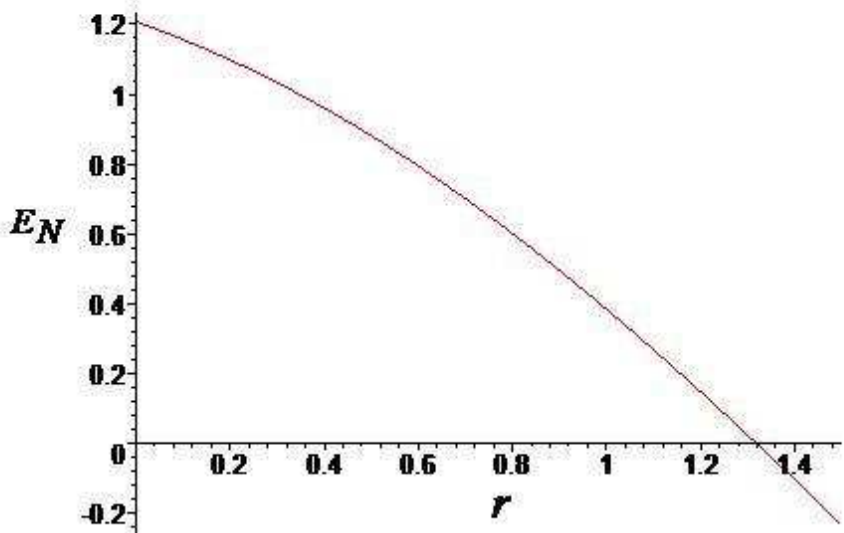
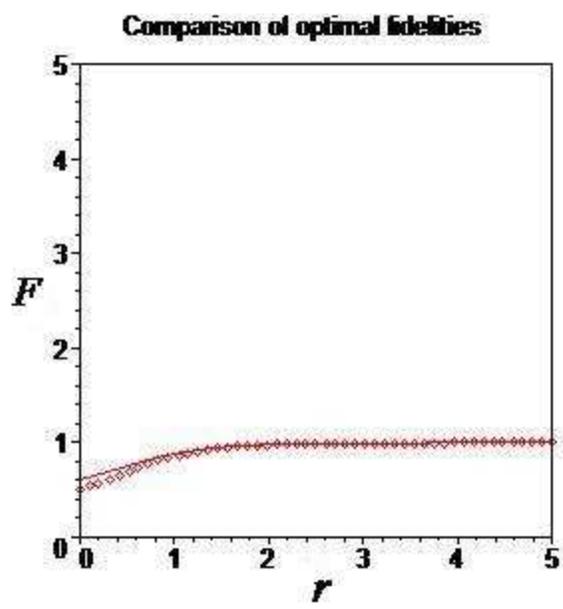
\end{document}